\documentclass[
    ,final            
  ]
  {aipproc}

\layoutstyle{8x11single}
\usepackage{amssymb}
\usepackage{amsmath}
\begin{document}

\title{Superfluid state of magnetoexcitons in double layer graphene structures}


\author{D. V. Fil, L. Yu. Kravchenko}
 {address={Institute for Single Crystals National Academy of Sciences of
Ukraine,\\ Lenin av. 60, 61001 Kharkov, Ukraine}}

\begin{abstract}
 The possibility of realization of a superfluid state of bound
electron-hole pairs (magnetoexcitons) with spatially separated
components in a graphene double layer structure (two graphene
layers separated by a dielectric layer) subjected by a strong
perpendicular to the layers magnetic field is analyzed.
 We show that the superfluid
state of magnetoexcitons may emerge only under certain imbalance
of filling factors
 of the layers.  The imbalance can be created by an
electrostatic field (external gate voltage). The spectrum of
elementary excitations is found and the dependence of the
Berezinskii-Kosterlitz-Thouless transition temperature on the
interlayer distance is obtained. The advantages of use graphene
double layer systems instead of double quantum well GaAs
heterostructures are discussed.
\end{abstract}

\maketitle


\section{Introduction}
\begin{figure}
  \includegraphics[height=.25\textheight]{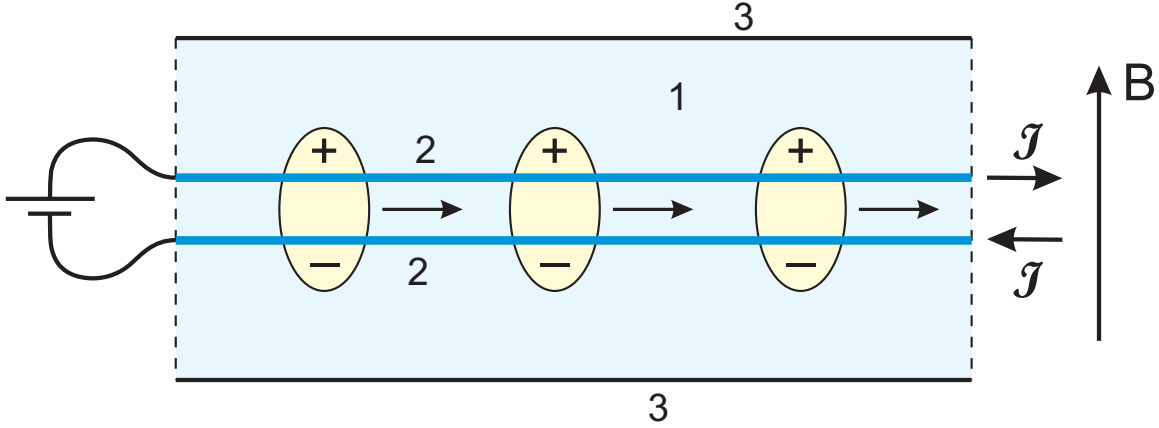}
  \caption{Bilayer graphene system (a dielectric matrix (1) with two
  embedded   graphene layers (2) inside a capacitor (3))
  that transmits the current by magnetoexcitons.
  }
  \label{f1}
\end{figure}

It is believed that excitons   may demonstrate superfluid
behavior. In bilayer systems superfluid excitons consisting of an
electron from one layer and a hole from the other layer behave as
superconductive ones. Indeed, having separate contacts in each
layer one can use excitons for a nondissipative transmission of an
electrical current from the source to the load (Fig. \ref{f1}).
The effect can be realized in double quantum wells in GaAs
heterostuctures \cite{1}. If such a system is subjected by a
strong perpendicular to the layers magnetic field and the total
filling factor of the Landau levels $\nu=1$,  electrons that
occupy quantum states in the zeroth Landau level in one layer
couple with holes (empty states in the zeroth Landau level) in the
other layer. Such pairs, called magnetoexcitons, are stable one,
in difference with optically excited indirect excitons in double
quantum wells \cite{2,3}. Stable excitons can emerge also in
bilayers made of \textit{n}-type and \textit{p}-type
two-dimensional conductors \cite{4,5}. In the latter systems the
nesting of the Fermi surfaces of electrons and holes is required
for the Bardin-Cooper-Schrieffer (BCS) pairing of \textit{n} and
\textit{p} carriers.

The discovery of graphene \cite{6,7}  has risen the idea of use
double layer graphene systems for the realization of the exciton
superconductivity. This question was already investigated in a
number of papers \cite{8,9,10,11,12}.

Graphene can be considered as a semiconductor with zero band gap.
Electron energy spectrum of graphene contains two Dirac points
that separate the electron and the hole subband. In a bilayer
structure the Fermi levels of the layers can be adjusted
independently by the gate voltage. If the Fermi level is in the
conduction band of one layer and is in the valence band of the
other layer we have a \textit{n-p} bilayer. The electron-hole
symmetry near the Dirac points ensures perfect nesting between the
electron and the hole Fermi surfaces. Such a situation was
considered in \cite {8,9, 10}. While the estimates in which the
screening effects are neglected \cite{8,9} yield very high BCS
temperature (hundreds of Kelvins), the screening may reduce the
critical temperature down to 6 orders \cite{10}.

In \cite{11,12} the superfluid transition in the rarefied gas of
magnetoexcitons was considered. Rarefied means that the number of
magnetoexcitons is much smaller than the number of states in the
Landau level. Since the critical temperature in 2D is proportional
to the concentration of the carriers, the rarefied gas of
magnetoexcitons cannot demonstrate record critical parameters. The
aim of this paper is to analyze the case of high magnetoexciton
density in bilayer graphene systems. Here we do not consider the
effect of screening. In quantum Hall systems screening is expected
to be small due to  finite gaps between Landau levels.

\section{Interlayer phase coherence in a double layer graphene system}

In conventional quantum Hall bilayers the case of high
magnetoexciton density is realized  at zero imbalance of filling
factors of the layers ($\nu_1=\nu_2=1/2$). The system at zero or
moderate imbalance of filling factors can be regarded as an
easy-plane quantum ferromagnet \cite{13}. Here we extend the
approach of  Ref. \cite{13} to a graphene bilayer system.

The Landau level energies and eigenfunctions for graphene are
given by the equation \cite{14}
\begin{equation}\label{1}
 - i \hbar v_F \left(%
\begin{array}{cc}
  0 & \frac{\partial}{\partial x}+\frac{\alpha  x}{\ell^2} -i \alpha \frac{\partial}{\partial y}\\
  \frac{\partial}{\partial x}-\frac{\alpha  x}{\ell^2} +i \alpha \frac{\partial}{\partial y} & 0 \\
\end{array}%
\right)
\left(%
\begin{array}{c}
  \Psi_A \\
  \Psi_B \\
\end{array}%
\right)= E \left(%
\begin{array}{c}
  \Psi_A \\
  \Psi_B \\
\end{array}%
\right).
\end{equation}
Here $v_F\approx 10^6$  m/s is the Fermi velocity (for the Dirac
spectrum $v_F$ is the matter parameter independent of the Fermi
energy), $\ell=\sqrt{\hbar c/eB}$ is the magnetic length,
$\alpha=\pm 1$ is the valley index (that corresponds to the
carriers with the pseudo-momentum close to the ${\bf
K}=(2\pi/3a,2\pi/3\sqrt{3}a)$ and ${\bf
K}'=(2\pi/3a,-2\pi/3\sqrt{3}a)$ Dirac points, correspondingly),
the spinor index $A(B)$ corresponds to the $A(B)$ sublattice of
graphene. The solution of Eq. (\ref{1}) yields the energies of the
Landau levels in graphene
\begin{equation}\label{2}
    E_0=0,\quad E_{\pm N}= \pm \frac{\hbar v_F}{\ell}\sqrt{2 N},
\end{equation}
and the corresponding eigenfunctions $\Psi_{N,\alpha}$
\begin{equation}\label{3}
  { \Psi}_{0,-1}(x,y)=\frac{e^{-i k
y}}{\pi^{1/4}\sqrt{\ell L_y}}e^{-\frac{(x-X)^2}{2\ell^2}}\left(%
\begin{array}{c}
  1 \\
  0 \\
\end{array}%
\right), \quad {\Psi}_{0,+1}(x,y)=\frac{e^{-i k
y}}{\pi^{1/4}\sqrt{\ell L_y}}e^{-\frac{(x-X)^2}{2\ell^2}}\left(%
\begin{array}{c}
  0 \\
  1 \\
\end{array}%
\right),
\end{equation}
\begin{eqnarray}\label{4}
  {\Psi}_{\pm N,-1}(x,y)=\frac{e^{-i k
y}}{\pi^{1/4}\sqrt{\ell L_y}\sqrt{2^{N+1}N!}}e^{-\frac{(x-X)^2}{2\ell^2}}\left(%
\begin{array}{c}
  \mp H_N\left(\frac{x-X}{\ell}\right) \\
  i \sqrt{2N} H_{N-1}\left(\frac{x-X}{\ell}\right)\\
\end{array}%
\right), \cr {\Psi}_{\pm N,+1}(x,y)=\frac{e^{-i k
y}}{\pi^{1/4}\sqrt{\ell L_y}\sqrt{2^{N+1}N!}}e^{-\frac{(x-X)^2}{2\ell^2}}\left(%
\begin{array}{c}
  i \sqrt{2N} H_{N-1}\left(\frac{x-X}{\ell}\right) \\
  \mp H_N\left(\frac{x-X}{\ell}\right)\\
\end{array}%
\right).
\end{eqnarray}
Here $X=k\ell^2$ is the guiding center of the orbit,  $H_N(x)$ is
the Hermite polynomial, and $L_y$ is the size of the system in $y$
direction. Approximation (1) is valid for $|E_N|\ll 3 t$, where
$3t\approx 8$ eV is the half-width of the energy band of graphene
(in the tight-binding approximation with the nearest-neighbor
hopping energy $t$). We do not take into account the Zeeman
splitting because it is much smaller than the energy distance
between the Landau levels.

Due to the spin and valley degeneracy the total number of quantum
states in the Landau level is $4S/2\pi \ell^2$ ($S$ is the area of
the system). We define the filling factor of the Landau level as
$\nu=2 \pi \ell^2 n_f$, where $n_f$ is the concentration of
occupied states. A completely filled level in graphene has the
filling factor $\nu=4$. In undoped graphene the negative levels
are completely filled, the positive levels are empty and the
zeroth level has the filling factor $\nu=2$.

The electron-hole pairing occurs between the carriers that belong
to a partially filled Landau level (we will call it the active
level). In particular, the pairing of electrons and holes
belonging to the zeroth Landau level may take place.  The
many-body wave function that describes the state with such a
pairing can be written analogously to \cite{13}
\begin{equation}\label{5}
 |\Psi\rangle =
 \prod_{X}\prod_{\beta}\left(\cos\frac{\theta_{\beta}}{2}
 a^+_{1,X\beta}+ e^{i \varphi_\beta}\sin \frac{\theta_\beta}{2}
 a^+_{2,X\beta}\right)|0\rangle.
\end{equation}
Here $\beta=(\alpha,\sigma)$ are the sets of valley and spin
quantum numbers (below we will notate them by digits
$\beta=1,2,3,4$), $a^+_{i,k\beta}$ is the operator of creation of
an electron in the $i$-th  layer in the active Landau level,
$\varphi_\beta$ is the phase of the order parameter for the
electron-hole pairing $\Delta_\beta=\langle\Psi
|a_{1,\beta}^+a_{2,\beta}|\Psi\rangle=(1/2)\sin \theta_\beta
e^{i\varphi_\beta}$, the parameter $\theta_\beta$ is connected
with the filling factors of the component $\beta$ by the relation
$\nu_{1(2),\beta}=(1\pm\cos\theta_\beta)/2$. The vacuum state
$|0\rangle$ is the state with empty zeroth (active) and higher
Landau levels in both layers.

Note that the function (\ref{5}) can be presented in another form
\begin{equation}\label{5-1}
 |\Psi\rangle =
 \prod_{X}\prod_{\beta}\left(\cos\frac{\theta_{\beta}}{2}
 + e^{i \varphi_\beta}\sin \frac{\theta_\beta}{2}
 a^+_{2,X\beta}h^+_{1,X\beta}\right)|vac\rangle,
\end{equation}
where $h^+_{i,X\beta}$ is the hole creation operator, and the
vacuum state $|vac\rangle$ is the state, where the zeroth Landau
level in the layer 2 is empty, while in the layer 1 this level is
completely filled. From (\ref{5-1}) it becomes clear that the
function (\ref{5}) is just the analog of the BCS wave function.
The state (\ref{5}) is usually called the state with a spontaneous
interlayer phase coherence.

We imply that the Coulomb energy $E_c=e^2/\varepsilon \ell$
($\varepsilon$ is the dielectric constant of the matrix in which
the graphene layers are embedded) is much smaller than the energy
distance between the active and the nearest passive (completely
filled or completely empty) Landau level and take into account
only the Coulomb interaction between electrons in the active
level. Below we consider the case of $N=0$ active level. The
Coulomb interaction has the form
\begin{equation}\label{6}
    H_C=\frac{1}{2S}\sum_{i,i'}\sum_{\bf q}
    V_{i,i'}(q)\hat{\rho}_i({\bf q})\hat{\rho}_{i'}(-{\bf q}),
\end{equation}
where $V_{i,i'}(q)=(2\pi e^2/\varepsilon q) \exp(-q d|i-i'|)$ is
the Fourier component of the Coulomb potential, $d$ is the
interlayer distance, and
\begin{equation}\label{7}
\hat{\rho}_i({\bf q})=\sum_{\beta} \sum_k a^+_{i,k+q_y/2,\beta}
a_{i,k-q_y/2,\beta} \exp\left(-iq_x k
\ell^2-\frac{q^2\ell^2}{4}\right)
\end{equation}
is the Fourier component of the electron density.

We consider the bilayer system situated inside the capacitor that
creates an electrostatic field normal to the graphene layers.
Varying the electrostatic field one can change the imbalance of
filling factors of the layers. The energy of the system in the
state (\ref{5}) reads as
\begin{equation}\label{8}
    E=\frac{S}{2\pi \ell^2}\left(W
    \left(\sum_\beta\tilde{\nu}_\beta\right)^2
    -\mathcal{J}_0\left(1+\sum_\beta\tilde{\nu}_\beta^2\right)-\mathcal{J}_1
    \left(1-\sum_\beta\tilde{\nu}_\beta^2\right)-  eV
    \sum_\beta\tilde{\nu}_\beta\right),
\end{equation}
where
$$
W=\frac{e^2 d}{\varepsilon \ell^2}, \quad
\mathcal{J}_0=\sqrt{\frac{\pi}{2}}\frac{e^2}{\varepsilon\ell},
\quad
\mathcal{J}_1=\sqrt{\frac{\pi}{2}}\frac{e^2}{\varepsilon\ell}
\exp\left(\frac{d^2}{2\ell^2}\right){\rm
erfc}\left(\frac{d}{\sqrt{2}\ell}\right)
$$
are the energy constants that describe the direct interlayer
interaction, the exchange intralayer interaction and the exchange
interlayer interaction (${\rm erfc}(x)$ is the complementary error
function). In (\ref{8}) $V$ is the external gate voltage caused by
the capacitor, and
$\tilde{\nu}_\beta=(\nu_{1,\beta}-\nu_{2,\beta})/2$ is the filling
factor imbalance for the component $\beta$
($|\tilde{\nu}_\beta|\leq 1/2$). In  (\ref{8})  the interaction of
electrons with the positively charged background is included. The
result (\ref{8}) can also be obtained in the standard mean-field
approach \cite{15-1}.

The interaction constants satisfy the inequalities
$\mathcal{J}_0>\mathcal{J}_1$ and
$W-\mathcal{J}_0+\mathcal{J}_1>0$. One finds that at $V=0$ the
minimum of (\ref{8}) corresponds to
$\tilde{\nu}_1=\tilde{\nu}_2=1/2$,
$\tilde{\nu}_3=\tilde{\nu}_4=-1/2$. At such $\tilde{\nu}_\beta$
all the order parameters $\Delta_\beta$ are equal to zero and
there is no electron-hole pairing. If the gate voltage is in the
range
\begin{equation}\label{9}
    \mathcal{J}_0-\mathcal{J}_1<eV<2W-\mathcal{J}_0+\mathcal{J}_1,
\end{equation}
the minimum is reached for
\begin{equation}\label{10}
    \tilde{\nu}_1=\tilde{\nu}_2=\frac{1}{2},\quad
    \tilde{\nu}_3=\frac{eV-W}{2(W-\mathcal{J}_0+\mathcal{J}_1)}, \quad
    \tilde{\nu}_4=-\frac{1}{2},
\end{equation}
and $\Delta_3\ne 0$.

For
\begin{equation}\label{11}
     2W+\mathcal{J}_0-\mathcal{J}_1\leq e V \leq 4W-\mathcal{J}_0+\mathcal{J}_1
\end{equation}
the energy minimum corresponds to
\begin{equation}\label{12}
    \tilde{\nu}_1=\tilde{\nu}_2=\tilde{\nu}_3=\frac{1}{2},\quad
    \tilde{\nu}_4=\frac{eV-3W}{2(W-\mathcal{J}_0+\mathcal{J}_1)},
\end{equation}
and $\Delta_4\ne 0$. Thus we conclude that an imbalance of filling
factors is required for the electron-hole pairing and only
electrons and holes with one $\beta$ are involved in the pairing.
In the special cases $eV=W$ and $eV=3W$ the imbalance of the
active component has zero value, and the order parameter for the
electron-hole pairing reaches the highest value.

The difference between the case of a quantum Hall bilayer in GaAs
with the total filling factor $\nu=1$ and the case of the graphene
bilayer system is the following. There is only one component in
the first case and an imbalance of that component increases the
direct interaction energy. In the second case positive imbalance
of some components can be compensated be negative imbalance of the
other components. The latter situation is similar to one that
takes place in $\nu=2$ quantum Hall bilayers \cite{15}.

\section{Critical temperature}

The state with moving superfluid pairs is described by the order
parameter with a spatially dependent phase. The many-body wave
function of the state with a constant superfluid magnetoexciton
current in $x$ direction reads as
\begin{equation}\label{13}
 |\Psi\rangle =
 \prod_{X}\left(\cos\frac{\theta_0}{2}
 a^+_{1,X\beta}+ e^{i Q X}\sin \frac{\theta_0}{2}
 a^+_{2,X\beta}\right)|0\rangle
\end{equation}
(the only part that corresponds to the active component is
displayed). The part of energy that depends on $\theta_0$ and $Q$
is equal to
\begin{equation} \label{14}
    E_\textrm{mf} = -\frac{e \tilde{V} S \cos \theta_0}{2\pi\ell^2}+
    \frac {S} {4 \pi \ell^2} \Bigg(\frac{W-\mathcal{J}_0}{2}
    \cos^2 \theta_0 -  {F} _{D} (Q) \sin^2\theta_0 \Bigg),
\end{equation}
where
\begin{equation} \label{15}
     {F} _{D} (q) = \frac {e^2} {2\varepsilon \ell} \int_0 ^ {\infty}
    d k \; e ^ {-\frac {k^2
    } {2}} J_0 (k q \ell) \; e ^ {-k d/\ell}
\end{equation}
($J_0(q)$ is the Bessel function). At small $Q$ Eq. (\ref{14}) is
reduced to $E_\textrm{mf}=S(const+\rho_{s0} Q^2/2)$, where
\begin{equation}\label{16}
\rho_{s0}= \sin^2 \theta_0 \frac{e^2}{16 \pi \varepsilon
\ell}\left[
\sqrt{\frac{\pi}{2}}\exp\left(\frac{d^2}{2\ell^2}\right){\rm erfc}
   \left(\frac{d}{\sqrt{2}\ell}\right)\left(1+\frac{d^2}{\ell^2}\right)-\frac{d}{\ell}\right]
\end{equation}
is the mean-field value of the superfluid stiffness (we take into
account that $Q=\nabla \varphi$). At finite temperatures the
excitations reduce this quantity. Therefore, the superfluid
stiffness $\rho_s$ depends on temperature.

Since the gas of electron-hole pairs in  bilayers is a
two-dimensional one, the transition into the superfluid state is
the Berezinskii-Kosterlitz-Thouless (BKT) transition. The critical
temperature of the BKT transition is given by the equation
\begin{equation}\label{17}
    T_c=\frac{\pi}{2}\rho_s(T_c).
\end{equation}
To obtain the  dependence $\rho_s(T)$ one should find the spectrum
of elementary excitations. Extending the approach \cite{16,17} to
the general case of an arbitrary angle between the wave vector
$\mathbf{q}$ and $\nabla\varphi$ we arrive at the following
expression for the energy of excitations
\begin{equation} \label{18}
E ({\bf q}) = \sqrt {\epsilon_{{\bf q},Q}  (\epsilon_{{\bf q},Q}
+2 \gamma_{{\bf q},Q}\sin^2 \theta_0)} + \cos \theta_0 v_{{\bf
q},Q},
\end{equation}
where
\begin{equation} \label{19} \epsilon_{{\bf q},Q} =2 {F} _{D}
(Q) -  {F} _{D} (|{\bf q}+Q \hat{x}|) -  {F} _{D} (|{\bf q}- Q
\hat{x}|),
\end{equation}
\begin{equation}\label{20}
v_{{\bf q},Q}= {F} _{D} (|{\bf q}+Q\hat{x}|) -  {F}_{D} (|{\bf q}-
Q\hat{x}|),
\end{equation}
\begin{equation} \label{21}
\gamma_{{\bf q},Q}  =  \left [ {H} ({\bf q},Q) - {F} _{S} (q) +
\frac{{F} _{D} (|{\bf q}+ Q \hat{x}|)+ {F} _{D} (|{\bf q}- Q
\hat{x}|)}{2}\right]
\end{equation}
with
\begin{equation}\label{22}
\mathcal {H} ({\bf q},Q)=\frac {e^2} {2 \varepsilon \ell^2} \; e ^
{-\frac {q^2 \ell^2} {2}}
    \frac {1 - e ^ {-d q }\cos(q_y Q \ell^2)} {q },
 \end{equation}
and
\begin{equation} \label{23}
     {F} _{S} (q) = \frac {e^2} {2\varepsilon \ell} \int_0 ^ {\infty}
    d k \; e ^ {-\frac {k^2
    } {2}} J_0 (k q \ell)
\end{equation}
($\hat{x}$ is the unit vector in $x$ direction). Having the
spectrum of elementary excitations one can compute the free energy
\begin{equation}\label{24}
    F=E_\textrm{mf}+T\sum_{{\bf q}} \ln
    \left(1-\exp\left(-\frac{E({\bf q})}{T}\right)\right) \
\end{equation}
and the density of the superfluid current
\begin{equation}\label{25}
 j_s=\frac{1}{S \hbar}
\frac{\partial F}{\partial Q}.
\end{equation}
On the other hand, at $Q\to 0$ the superfluid current is connected
with the superfluid stiffness by the relation
\begin{equation}\label{26}
 j_s=\frac{\rho_s(T)}{\hbar} Q.
\end{equation}
Using (\ref{23})-(\ref{25}), we obtain the following expression
for the superfluid stiffness
\begin{equation}\label{27}
 \rho_s(T)=\rho_{s0}+\frac{1}{S}\lim_{Q\to 0}\frac{1}{Q}\sum_{{\bf q}}\frac{\partial E({\bf
    q})}{\partial
    Q}N_B\left(E({\bf q})\right),
\end{equation}
where $N_B(E)=[\exp(E/T)-1]^{-1}$ is the Bose distribution
function. For the spectrum $E({\bf q})=E_0(q)+\hbar \mathbf{q}
\mathbf{v}$ (where $\mathbf{v}$ is the superfluid velocity, and
$E_0(q)$ is the spectrum at $\mathbf{v}=0$) the temperature part
of Eq. (\ref{27}) yields the standard expression for the normal
density \cite{18}. The present situation differs from \cite{18}
because the dependence of the spectrum on the superfluid velocity
$\mathbf{v}=\hbar Q/M$ ($M=\hbar^2\sin^2\theta_0/8\pi \rho_{s0}
\ell^2$ is the magnetic mass of the electron-hole pair) is more
complicate.

The nature of such a difference is the following. In certain
sense, superfluidity of electron-hole pairs in the bilayer can be
considered as a kind of a counterflow superfluidity \cite{19}.
Indeed, due to the Coulomb interaction electrons from the top
layer are coupled with holes from the bottom layer as well as
holes from the top layer are coupled with electrons from bottom
layer. In other words, one can say about two species of bosons of
different polarization. In such a two-specie system only a
counterflow motion of species is possible. In the limit of low
mangetoexciton density the second specie can be considered as an
inert background and the system behaves as a one-component gas of
interacting Bose particles. But at zero imbalance of the active
component we have two equivalent species that move in opposite
directions with the same velocities. In particular, it results in
that the last term in (\ref{18}) goes to zero at $\theta_0=\pi/2$.
The counterflow superfluidity is a special kind of the
two-component superfluidity. In similarity with the spectrum
(\ref{18}), the spectrum of elementary excitations of
two-component superfluid system demonstrates a complicate
dependence on the superfluid velocities (see, for instance,
\cite{20}).

Let us describe the behavior of the spectrum (\ref{18}) at
different $d/\ell$ and $Q=0$. At $d\to 0$ the spectrum reduces to
the quadratic one $E(q)=\hbar^2 q^2/2 M$ (for small $q$).
According to the Landau criterium of superfluidity it means the
absence of superfluidity. At  $d\approx\ell$ the roton-like
minimum emerges at the dependence $E(q)$. The depth of this
minimum increases under the increase of $d/\ell$ and at some
critical $d=d_c$ the minimum touches the $q$ axis. Such a behavior
of the spectrum means that at $d>d_c$ the system becomes instable
with respect to a formation of the charge density wave. We imply
the superfluidity does not survive in the charge density wave
state. At $\theta_0=\pi/2$ (that corresponds to zero imbalance of
the active component) $d_c\approx 1.175 \ell$. One can expect that
at $d\to 0$ and $d\to d_c$ the superfluid stiffness is suppressed,
that results in a lowering of the critical temperature.

\begin{figure}
  \includegraphics[height=.3\textheight]{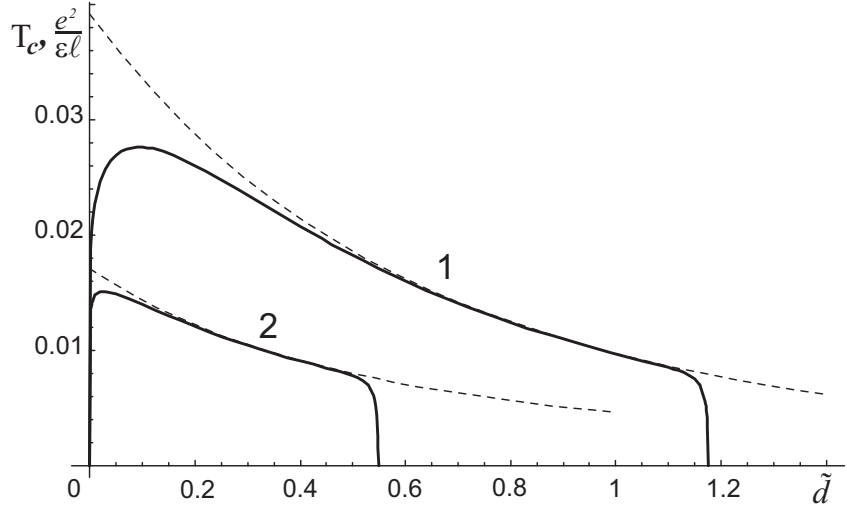}
  \caption{The dependence of the critical temperature
  on the interlayer distance ($\tilde{d}=d/\ell$); 1 - coupling in the zeroth Landau level,
  2- coupling in the $N=\pm 1$ Landau levels.
  The quantities $\pi \rho_{s0}/2$ are shown by dashed lines.} \label{f2}
\end{figure}

The critical temperature obtained from Eq. (\ref{17}) for the case
of zero imbalance ($\theta_0=\pi/2$) is shown in Fig. \ref{f2}.
One can see  that, indeed, at $d\to 0$ and $d\to d_c$ the critical
temperature goes to zero. We also find that at intermediate $d$
the critical temperature can be evaluated with a good accuracy
from the mean-field superfluid stiffness. Since the stiffness is
proportional to $\sin^2 \theta_0=1-4\tilde{\nu}^2$, the highest
critical temperatures can be reached at zero imbalance of the
active component (or at the filling factors of the zeroth Landau
level $\nu_1=3/2$, $\nu_2=5/2$ and  $\nu_1=1/2$, $\nu_2=7/2$).

The approach presented can be easily generalized for the case
where electrons in the $+N$ Landau level are coupled with holes in
the $-N$ level. This situation can be realized at large external
gate voltage $eV>2\sqrt{2} \hbar v_F/\ell$. To obtain the critical
temperature in the latter case one should take into account the
additional factor
$f_N(q\ell)=[L_N(q^2\ell^2/2)+L_{N-1}(q^2\ell^2/2)]/2$ in the
electron density operator (\ref{7})  ($L_N(x)$ are the Laguerre
polynomials). Respectively, the factor $f_N^2(k)$ appears under
the integrals in (\ref{15}) and (\ref{23}), and the factor
$f_N^2(q\ell)$ should be added into the definition (\ref{22}).
Here we consider the case of the $\pm 1$ active level. The
mean-field superfluid stiffness in this case reads as
\begin{equation}\label{28}
\rho_{s0}=\sin^2 \theta_0 \frac{e^2}{16 \pi \varepsilon
\ell}\left[\sqrt{\frac{\pi}{2}}e^{\frac{d^2}{2\ell^2}}{\rm erfc}
   \left(\frac{d}{\sqrt{2}\ell}\right)\frac{7+13 (d/\ell)^2+7(d/\ell)^4+(d/\ell)^6}{16}
   -\frac{d(3\ell^2+d^2)^2}{16\ell^3}\right].
\end{equation}
At all $d/\ell$ the superfluid stiffness for the $\pm 1$ Landau
levels is smaller than one for the zeroth level (at the same
imbalance of the active component). For the $\pm 1$ active levels
the critical interlayer distance is $d_c\approx 0.549 \ell$. Using
the procedure, described above, we compute the critical
temperature. The result is shown in Fig. \ref{f2}.

\section{Discussion}

We conclude that for the gas of magnetoexcitons in a bilayer
graphene system the maximum critical temperature of the superfluid
transition  can be achieved for the coupling of electrons and
holes in the zeroth Landau level and at special values of the
imbalance of filling factors of the zeroth level ($\nu_1-\nu_2=\pm
1$ and $\nu_1-\nu_2=\pm 3$).

For the formation of magnetoexcitons the Coulomb energy should be
much smaller than the energy distance between an active  and the
nearest passive level. The same condition allows to neglect the
effect of screening
 on the critical temperature. Indeed,
screening can be taken into account by a substitution a
$q$-dependent dielectric function $\varepsilon(q)$ instead of the
dielectric constant $\varepsilon$ into the Fourier components of
the Coulomb potential (see, for instance, \cite{my1}). In the case
considered here the difference $\varepsilon(q)-\varepsilon$ is
small by the parameter $E_c/(E_1-E_0)$.

For the graphene system the inequality $E_c\ll E_1-E_0$ is just
the condition on the dielectric constant $\varepsilon\gg
\varepsilon_c\approx 1.5$. At large $\varepsilon$ this inequality
is fulfilled at all values of magnetic fields (of course, it is
implied that the temperature is much smaller than $E_1-E_0$). The
situation differs significantly from one that takes place in
quantum Hall bilayers in GaAs heterostructures. For the carriers
with the quadratic dispersion the energy distance between the
Landau levels is proportional to B ($\omega_c=e B/m_* c$, where
$m_*$ is the effective mass) and the condition $E_c<\hbar
\omega_c$ is fulfilled only at  large magnetic fields ($B\gtrsim
10$ T for GaAs). The graphene bilayers require much lower $B$.

\begin{figure}
  \includegraphics[height=.3\textheight]{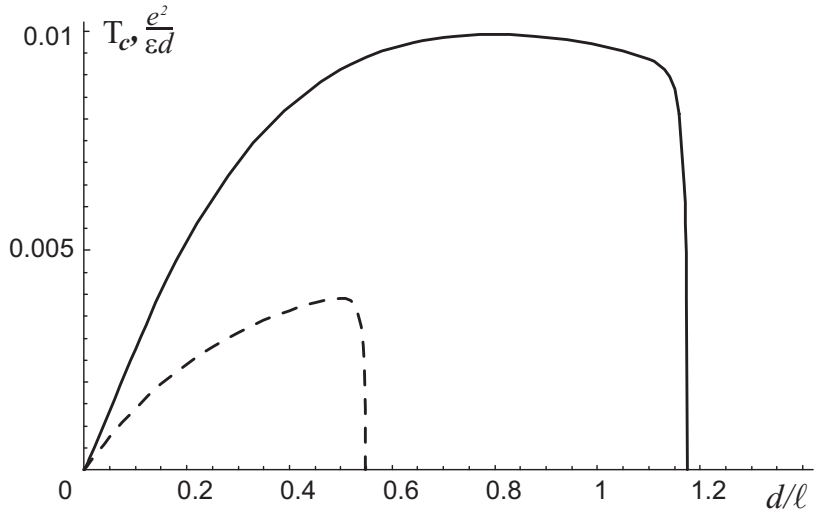}
  \caption{The dependence of the critical temperature
  on the inverse magnetic length at fixed $d$. Solid curve -
  magnetoexcitons in the zeroth Landau level, dashed curve -
  magnetoexcitons in the $N=\pm 1$ levels.
  } \label{f3}
\end{figure}

It is interesting to analyze how the critical temperature depends
on $B$. Let we have a bilayer system with a given interlayer
distance $d$. We vary magnetic field  adjusting simultaneously the
gate voltage (to keep the imbalance close to the optimal value)
and try to achieve the highest critical temperature. Presenting
the dependence of Fig. \ref{f2} in $e^2/\varepsilon d$ units we
obtain (Fig. \ref{f3}) the dependence of $T_c$ on $\ell^{-1}$ (or
on $B^{1/2}$). One can see that the critical temperature is
restricted from above by $T_{c,max}\approx 0.01 e^2/\varepsilon
d$, and the maximum is reached for $\ell$ in the interval $[d\div
2d]$. To achieve the critical temperature $T_c=1$ K we may take
the bilayer with the interlayer distance $d$ as large as 400 \AA \
(for $\varepsilon=3.9$ (SiO$_2$)) and apply comparatively low
magnetic fields $B\gtrsim 0.4$ T. At the same time, our estimate
shows that extremely high magnetic fields are required if we want
to realize a superfluid state of magnetoexcitons at high
temperatures. For instance, for $d\approx 10$ \AA \ we have
$T_{c,max}\approx 40 K$, but the magnetic field should be $B>600$
T.

There is another reason that makes the case of low magnetic field
important. Low fields mean large critical interlayer distances. As
was shown in \cite{21,22}, in the bilayer system the genuine
superconductivity is reached in the limit of zero interlayer
tunneling \footnote{Strictly speaking, dissipation in 2D
superfluid systems remains nonzero at all temperatures \cite{n1}.
In the presence of a finite superfluid velocity, vortex pairs can
unbind even below $T_c$. Although this  process leads to the decay
of any finite persistent current, the decay rate vanishes at zero
velocity below $T_c$.
  With reference to the quantum Hall bilayers it was discussed in
\cite{13,16}. At $T\ll T_c$ and for the current much lower than
the mean-field critical current \cite{16,17} the decay rate is
exponentially small and one can say about genuine
superconductivity} If the systems with a finite tunneling is used
for the transmission of the current from the source to the load
(Fig. \ref{f1}), the dissipation is nonzero. The dissipation is
connected with that the state becomes nonstationary at nonzero
difference of electrochemical potentials between the layers. This
difference is required to provide the current in the load circuit.
The power of losses is proportional to the square amplitude of the
interlayer tunneling \cite{21,22}. A partial solution of this
problem was proposed in \cite{23}. In the setup, considered in
\cite{23}, a stationary state with moving electron-hole pairs can
be realized, but, regrettable, the setup \cite{23} (and any other
setup based on the same idea \cite{22}) cannot be used for the
transmission the current from the source to the load. The
amplitude of tunneling decreases exponentially with the increase
of the interlayer distance, and for $d\approx 400$ \AA \ the
effect of tunneling can be completely neglected.

\end{document}